\documentclass[twocolumn,showpacs,pra]{revtex4}

\usepackage[colorlinks=true,citecolor=blue,linkcolor=magenta]{hyperref}

\usepackage{amsmath}
\usepackage{amssymb}
\usepackage{txfonts}
\usepackage{graphicx}
\usepackage{epstopdf}
\usepackage{dcolumn}
\usepackage{bm}
\begin{document}
\title{Two-photon blockade in a cascaded cavity-quantum-electrodynamics system}
\author{Qian Bin}
\author{Xin-You L\"{u}}\email{xinyoulu@hust.edu.cn}
\author{Shang-Wu Bin}
\author{Ying Wu}\email{yingwu2@126.com}
\affiliation{School of Physics, Huazhong University of Science and Technology, Wuhan, 430074, P. R. China}
\date{\today}

\begin{abstract}
We investigate theoretically the model of a cavity-quantum-electrodynamics (QED) system that consists of two two-level atoms coupled to a single-mode cavity in the weak coupling regime, where the system is driven by quantum light. The dynamics behavior of the entire system is tackled in the framework of a cascaded quantum system. We find that the two-photon blockade with two-photon bunching and three-photon antibunching can be obtained even when the strong system dissipation is included. This result shows that our work has potential for realizing entangled photon pairs in a weakly coupled cavity. Moreover, we also analyze the photon statistics of the system in the case of out-of-resonance coupling between cavity and two nonidentical atoms. Here, an unconventional photon blockade effect with the suppression of two-photon correlation and enhancement of three-photon correlation can be realized, which shows many quantum statistical characteristics of cavity QED system in weak coupling.
\end{abstract}

\maketitle
\section{Introduction}
The realization of photon-photon interaction has been an attractive topic in quantum optics and quantum information processing for the past few decades. The key requirement for realizing this interaction is the generation and manipulation of photon, so we need to prepare a single-photon source. Photon blockade (PB) provides a way to realize this source. Here PB means that a single photon in a cavity can block the transmission of other photons as a result of the strong nonlinearity in the system~\cite{ref1}. Only when the first photon has left the system can the second photon enter it. Thus a sequence of single photons can be generated one by one and the system can be set as a single-photon source device, which is a useful application in modern quantum optics ~\cite{ref2-1,ref2-2,ref2-5}. The PB effect shows antibunching with the second-order photon correlation function and the sub-Poisson photon number statistics. Obviously, the PB realizes the change from the input of a classical light field to that of a non-classical light field. Moreover, the PB is also called a nonlinear quantum scissor, which can occur in a nonlinear system induced by the strong interaction between a cavity and two-level system~\cite{ref2-3,ref2-4}. The PB effect was first observed in a system that consisted of a single trapped atom coupled to an optical cavity~\cite{ref2-1}. This effect has been theoretically proposed in cavity QED, circuit QED and optomechanical systems~\cite{ref2-2,ref9,ref14,ref15,ref8,ref8-1,ref8-2,ref8-3}. It has also been experimentally demonstrated in a system that consists of a photonic crystal cavity strongly coupled to a quantum dot~\cite{ref16}. Furthermore, the PB effect has also been studied in other systems and knowledge of it has progressed enormously in recent years~\cite{ref3,ref3-1,ref3-2,ref4,ref5,ref6,ref6-2,ref6-3,ref6-4,ref11,ref11-1,ref12,ref13}.

Previously, studies on PB has been mainly observations of single-photon blockade effects. Recently, the study of PB has been extended to the multiphoton blockade. Here, there is a photon in cavity, but the second and third photons can still be absorbed via two-photon and three-photon processes, respectively. Such multiphoton blockade systems could generate multiphoton streams, leading to many applications in quantum nonlinear optics, like ideal entangled photon sources. To date, the multiphoton blockade has been studied in various configurations~\cite{ref7,ref17,ref17-1,ref18}. For example, it has been proposed that two-photon and three-photon blockades can be observed in the standard cavity QED in the past few years~\cite{ref7}. Here, the system consists of a Kerr nonlinearity in a cavity driven by a weak classical field. The reason for realizing the multiphoton blockade is the presence of strong nonlinear photon-photon interactions in the Kerr-type system. It has also been shown that the collective multiphoton blockade can be obtained in a cavity QED system that consists of a cavity strongly coupled to two atoms, where two atoms are driven by the coherent light field~\cite{ref20}. Recently, it has been shown that the two-photon blockade effect can be realized experimentally in an atom-driven cavity QED system~\cite{ref19}. The prerequisite for realizing these multiphoton blockades in these systems mentioned above is the presence of strong nonlinearities. However, it is still difficult to achieve strong nonlinearities in many systems, because the generation of strong nonlinearity needs strong nonlinear susceptibility, a strong coupling regime, or a high-quality cavity. One question that arises naturally is whether multiphoton blockade can be reached in a weak coupling regime or bad cavity system.

Motivated by the above question, in this paper we propose a system model that consists of two two-level atoms weakly coupled to a single-mode cavity. The system is a prototypical Tavis-Cummings (TC) model, called as target system, and which is driven by the output field of a source system. Here the source is made of a two-level atom driven by the classical light field~\cite{ref21}. The source and target systems could be operated in the cascaded quantum system. The output photons from the source system show antibunching with the second-order photon correlation function, so the output field of the source system is non-classical. We find that when the output field from the source is scanned into the TC model, the two-photon blockade with two-photon bunching and three-photon antibunching can occur in the target system. Here, because the system is coupled with two atoms, its nonlinearity is greater than that of a system with one coupling with an atom. However, we can still see that, with the decrease of the coupling strengths between the cavity and target two-level atoms, this two-photon blockade effect will disappear. Furthermore, we investigate the photon correlations when the cavity frequency is detuned with respect to two target two-level atoms. We can obtain the two-photon blockade in a weaker coupling regime. These results show that we can realize potentially the entangled photon pairs by using the weakly coupled cavity to convert an anti-bunched source modeled by the two-level system. We also observe the unconventional PB effect with the two-photon antibunching and three-photon bunching when the frequency detuning between target atoms is considered. The criterion for demonstrating the conditions of unconventional PB has been contested in a recent paper~\cite{ref21-1}. Moreover, we calculate numerically the photon number probabilities in the truncated Hilbert space, and show the distributions of the single-photon and two-photon with the absence of three-photon in the given parameter regimes.

Our paper is organized as follows. In Sec.\,II, we introduce the model and the ladder of the dressed states. In Sec.\,III, we demonstrate the existence of two-photon blockade by calculating numerically the second-order and third-order photon correlations of the cavity output field. Furthermore, we present the two-photon blockade in a weaker coupling regime when the frequency detunings between the cavity and target two-level atoms are included. We also show the unconventional PB effect when the atom-atom detuning is also considered. In Sec.\,IV, we give discussions for the applications of the results and the experimental realization of our proposal. In Sec.\,V, we give conclusions of our work.
\section{Model}
\begin{figure}
  \centering
  \includegraphics[width=8.5cm]{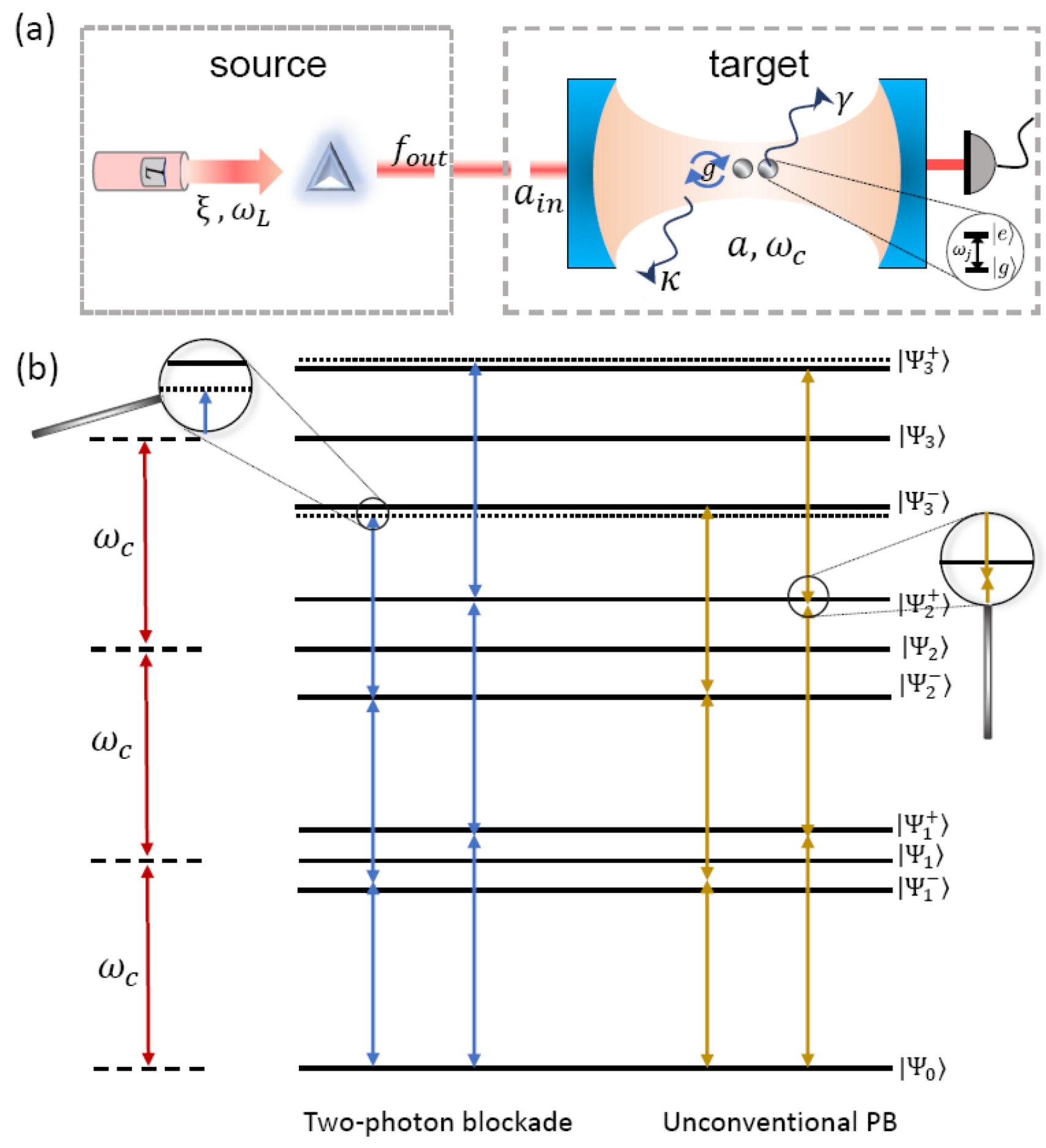}\\
  \caption{ (a) Schematic of the studied system. The cavity QED system consisting of two two-level atoms weakly coupled to a single-mode cavity driven by the emission of a quantum source. Here, the quantum source is made of a two-level atom driven by the classical light field, where $f_{out}$ is the output channel of the source system. $a_{in}$ is the input channel of the target. (b) Schematic energy-level structure illustrating the occurrence of the two-photon blockade and unconventional PB annotated at the bottom. The black solid lines represent the energy levels with the optical-state truncation.}\label{fig1}
\end{figure}

\begin{figure}
  \centering
  \includegraphics[width=8.5cm]{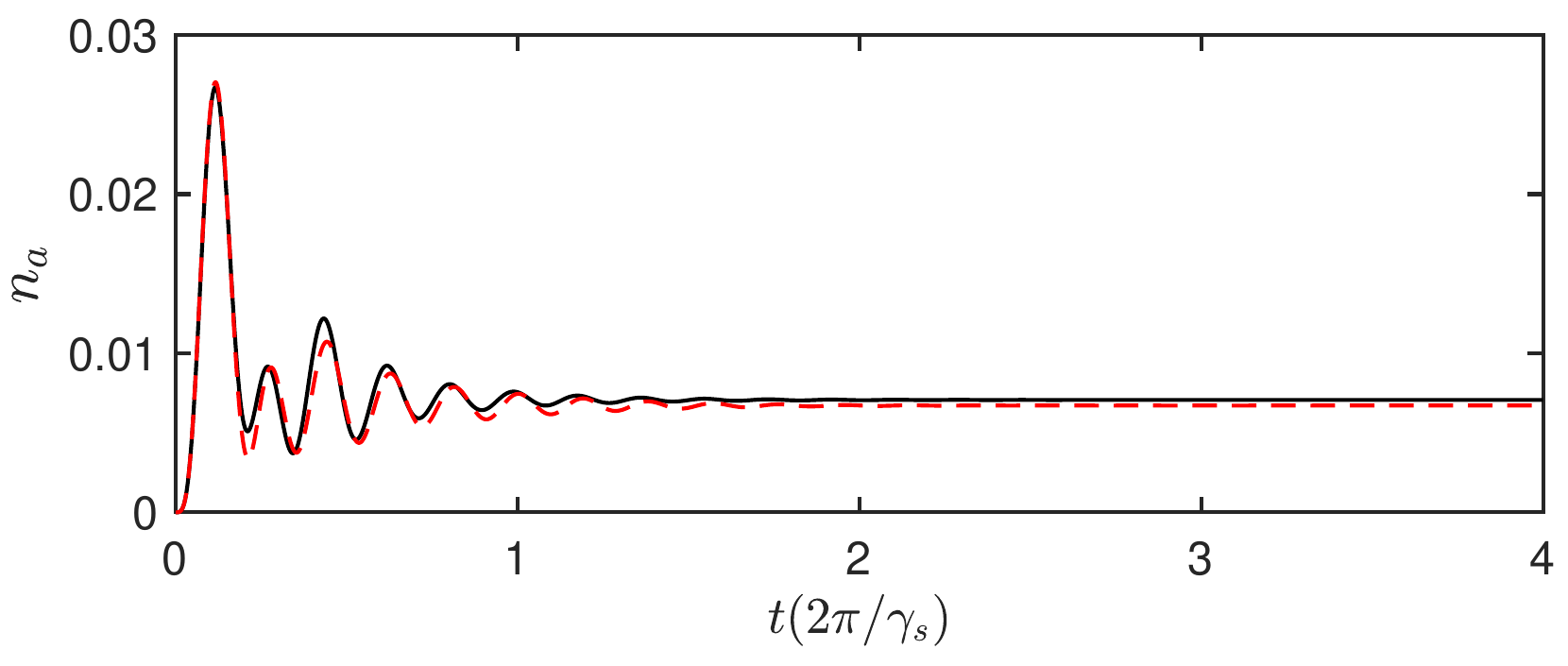}\\
  \caption{  Time evolution (in units of $2\pi/\gamma_s$) of the cavity mean photon number $n_a$ obtained by calculating Eq.~(\ref{eq3}) (black solid line) and Eq.~(\ref{eq6}) (red dashed line) when $\Delta_C/\gamma_s=5$. The system parameters used here are: $\xi/\gamma_s=1$, $g/\gamma_s=1.25$, $\kappa/\gamma_s=5$, $\gamma/\gamma_s=0.375$ and $\Delta_1=\Delta_2=\Delta_C$.}\label{fig2}
\end{figure}

As shown in Fig.~\ref{fig1}, we consider a cavity QED system that consists of two two-level atoms weakly coupled to a single-mode cavity driven by the quantum source. Here the source system is made of a two-level atom driven by a classical light field with frequency $\omega_L$ and drive amplitude $\xi$. We then consider transforming the source and target systems into a frame rotating with $\omega_L$ to remove the time dependence, so the source Hamiltonian can be given by $(\hbar=1)$
\begin{equation}\label{eq1}
H_s=\Delta_s\ \sigma_s^\dag \sigma_s+\xi \sigma_s+\xi^\ast \sigma_s^\dag,
\end{equation}
and for the target system, which has the prototypical Tavis-Cummings Hamiltonian~\cite{ref22}
\begin{align}\label{eq2}
H_{t}=&\Delta_C a^\dag a+\sum\limits_{j=1,2}\Delta_{j} \sigma_j^\dag \sigma_j
+\sum\limits_{j=1,2}g(\sigma_j^\dag a+\sigma_j a^\dag).
\end{align}
Here, $a$ $(a^\dag)$ is the annihilation (creation) operator of the cavity mode with resonant frequency $\omega_c$. $\sigma_s$ denotes the lowering operator of the source system with transition frequency $\omega_s$, $\sigma_j$ $(j=1,2)$ denotes the lowering operator of the target two-level atom with transition frequency $\omega_j$ $(j=1,2)$, and $g$ is the atom-cavity coupling strength. $\Delta_C=\omega_c-\omega_L$ and $\Delta_s=\omega_s-\omega_L$ are, respectively, the cavity and source frequency detunings with respect to the laser field. Here, the frequency of the source two-level atom is resonant with the cavity frequency, i.e, $\Delta_s=\Delta_C$. $\Delta_{j}=\omega_j-\omega_L$ is the detuning of the target atoms from the laser field. The target system is driven by the output field of the source system though the output channel $f_{in}$ and input channel $a_{in}$, and the main requisite is that the dynamics of the source system is not affected by the presence or absence of the target. We thus consider that there is a dissipative coupling between the source and target systems via a thermal bath, and we tackle the dynamics of the coupled system in the framework of a cascaded quantum system. The output field of the source is set as the input field of the target via equations of motion, and there is no back action from the target in the case. Under these conditions, we obtain the master equation~\cite{ref23,ref24,ref25,ref27,ref28,ref29,ref30,ref31}
\begin{align}\label{eq3}
\frac {d\rho}{dt}=&i[\rho,H_s+H_t]+\gamma_s\mathcal{L}[\sigma_s]+\kappa\mathcal{L}[a]
+\gamma\sum\limits_{j=1,2}\mathcal{L}[\sigma_j]\nonumber\\
&-\sqrt{\mu\gamma_s \kappa} \{[a^\dag,\sigma_s \rho]+[\rho \sigma_s^\dag,a]\},
\end{align}
with the superoperator $\mathcal{L}$ expressed as
\begin{equation}\label{eq4}
\mathcal{L}[\emph{O}]=\frac {1}{2}(2\emph{O}\rho\emph{O}^\dag-\rho\emph{O}^\dag\emph{O}-\emph{O}^\dag\emph{O}\rho).
\end{equation}
Here, $\gamma_s$ is the decay rate of the source; $\gamma$ and $\kappa$ are the decay rates of the target atoms and cavity, respectively. The parameter, $0\leq\mu\leq1$, includes the phenomena like back-scattering and absorption. $\sqrt{\mu\gamma_s \kappa}$ represents the dissipative coupling coefficient between the source and target systems ($\mu=1$). From the master equation~(\ref{eq3}), a stochastic Schr\"{o}dinger equation can be given for the cascaded system, where the evolution of stochastic wavefunction is expressed as
\begin{equation}\label{eq5}
\frac{d}{dt}|\Psi(t)\rangle=-i H_{\texttt{eff}}|\Psi(t)\rangle,
\end{equation}
where the non-Hermitian effective Hamiltonian can be written by~\cite{ref23,ref24,ref25,ref27}
\begin{align}\label{eq6}
H_{\texttt{eff}}=&(\Delta_s-i\frac {\gamma_s}{2})\ \sigma_s^\dag \sigma_s+(\Delta_C-i\frac {\kappa}{2}) a^\dag a+\sum\limits_{j=1,2}(\Delta_j-i\frac {\gamma}{2}) \sigma_j^\dag \sigma_j\nonumber\\
&+\sum\limits_{j=1,2}g(\sigma_j^\dag a+\sigma_j a^\dag)+\xi (\sigma_s+ \sigma_s^\dag)-i\sqrt{\gamma_s\kappa}a^\dag\sigma_s.
\end{align}
The last term of the above equation denotes that the transition of source system induces the generation of photons of the target system. It can be seen that the excitations are transferred from the source two-level emitter to the target cavity. To understand the dynamics of the system more clearly, in Fig.~\ref{fig2}, we display the comparison of the time evolutions of cavity mean photon numbers obtained by solving numerically Eq.~(\ref{eq3}) (black solid line) and Eq.~(\ref{eq6}) (red dashed line). We see that the two results are basically consistent. However, there is an obvious deviation between the higher-order correlation functions obtained by two methods. This is because in the process of solving an effective Hamiltonian, we neglect the influence of quantum jumps on the system. Therefore, in the next sections, we use the master equation~(\ref{eq3}) to calculate the dynamics of the system.
\section{Results for two-photon blockade and unconventional blockade}
\subsection{Two-photon blockade without atom-cavity detuning}
\begin{figure*}
  \centering
  \includegraphics[width=14.0cm]{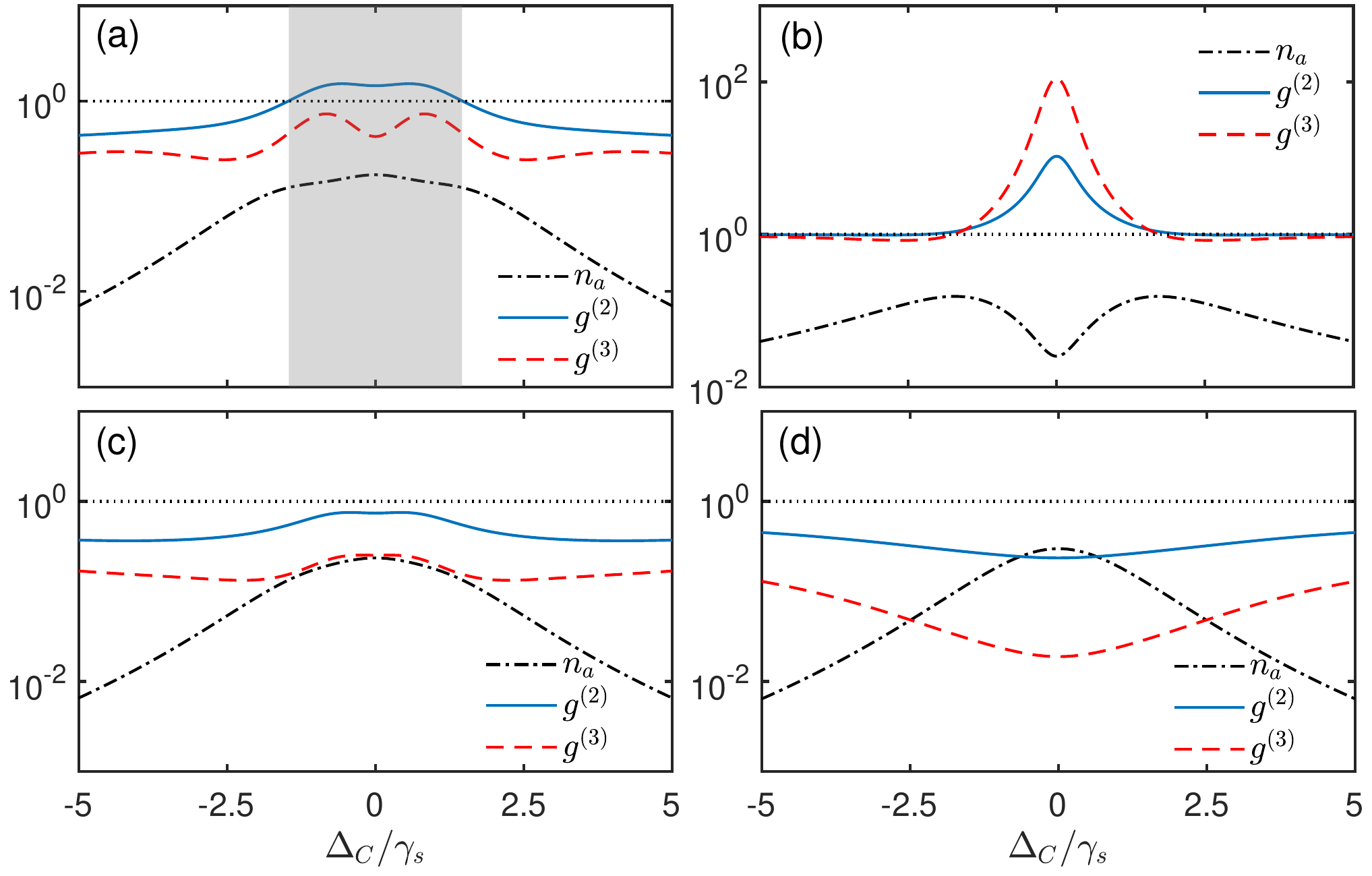}\\
  \caption{ The mean photon number $n_a$ (black dash-dotted curve), equal-time second-order photon correlation function $g^{(2)}$ (blue solid curve) and equal-time third-order photon correlation function $g^{(3)}$ (red dashed curve) of the output of cavity vs $\Delta_C$. Panels (a) and (b) correspond to the cavity QED system that consists of two atoms weakly connected to a single-mode cavity which is excited by quantum light and classical light, respectively; panels (c) and (d) correspond to the cavity QED system described by the Jaynes-Cummings (JC) model and empty cavity respectively, which are excited by the same quantum light. The system parameters used here are the same as in Fig.~\ref{fig2}.}\label{fig3}
\end{figure*}
\begin{figure}
  \centering
  \includegraphics[width=8cm]{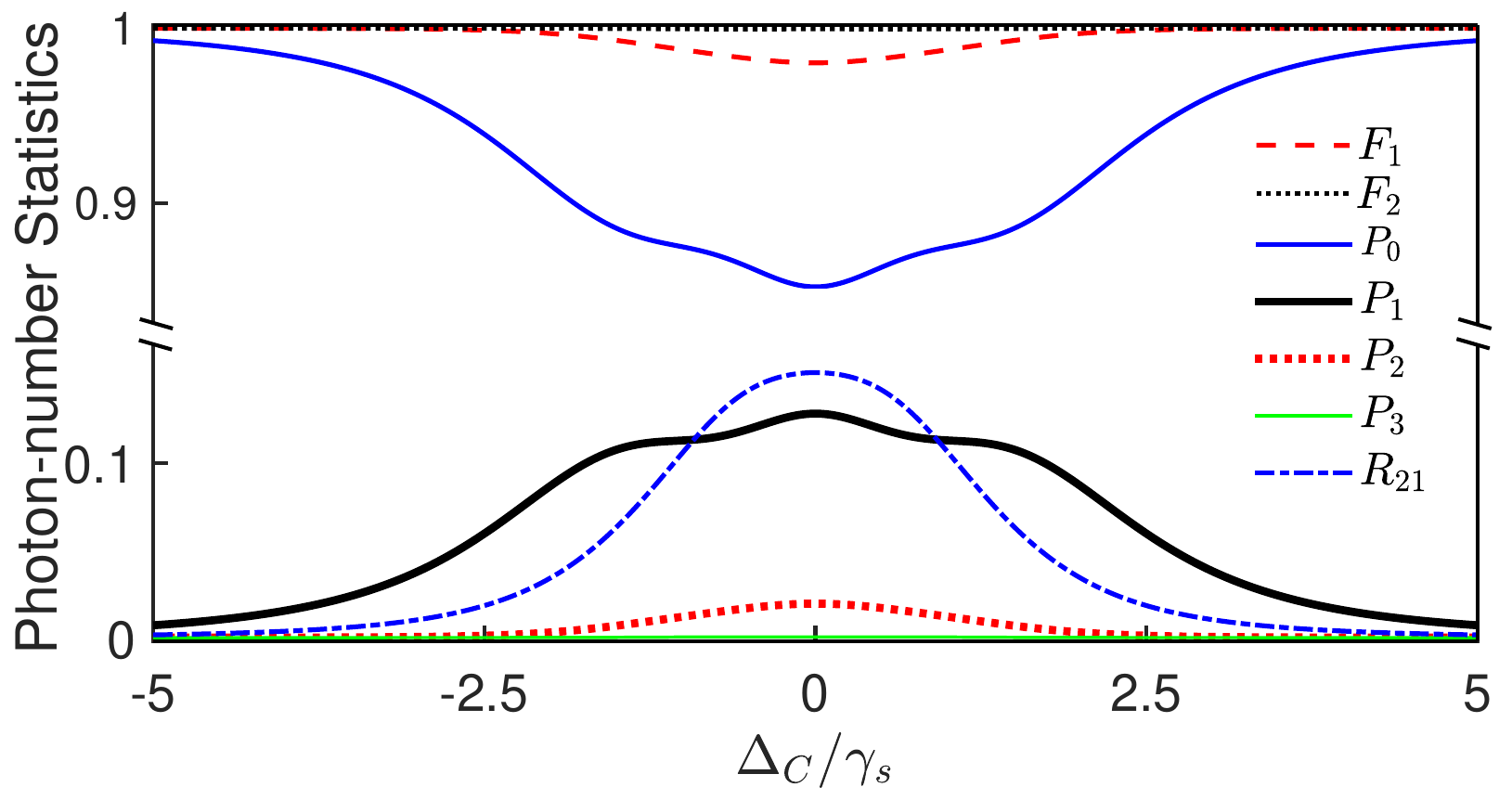}\\
  \caption{ Photon-number probabilities $P_n$, truncation fidelity $F_k$ and the ratio of two-photon state probability to one-photon state probability $R_{21}$ vs the cavity-field detuning $\Delta_C$. The system parameters used here are the same as in Fig.~\ref{fig2}}\label{fig4}
\end{figure}
We consider that our target system is designed to operate in a weak coupling regime, where the atom-cavity coupling strength is less than the dissipation of the system. In order to identify the photon-number statistics of the output field, we have to solve the photon correlation functions, which are defined in the steady-state limit as
\begin{gather}\label{eq7}
g^{(2)}(\tau)=\lim_{t\rightarrow\infty}\frac{\langle a^\dag(t)a^\dag(t+\tau)a(t+\tau)a(t)\rangle}{\langle a^\dag(t)a(t)\rangle^2},
\end{gather}
\begin{gather}\label{eq8}
g^{(3)}(\tau_1;\tau_2)=\lim_{t\rightarrow\infty}\frac{\langle a^\dag(t)a^\dag(t+\tau_1)a^\dag(t+\tau_2)a(t+\tau_2)a(t+\tau_1)a(t)\rangle}{\langle a^\dag(t)a(t)\rangle^3}.
\end{gather}
We assume that there is a zero delay between photons, i.e., $\tau=\tau_1=\tau_2=0$, and Eqs.~(\ref{eq7}) and~(\ref{eq8}) are expressed as
\begin{equation}\label{eq9}
g^{(n)}(0)=\frac{\langle a^{\dag n}a^n\rangle}{\langle a^\dag a\rangle^n}.
\end{equation}
Assuming $\omega_c=\omega_j$, i.e., $\Delta_C=\Delta_j$ $(j=1,2)$. Transforming it back to the laboratory frame, the eigenvalues and eigenstates of the TC Hamiltonian $H_t$ in the three-photon space can be given by
\begin{gather}\label{eq10}
E_1^{0}=\omega_c;~|\Psi_1^0\rangle=\frac {1}{\sqrt{2}}|0,g,e\rangle-\frac {1}{\sqrt{2}}|0,e,g\rangle,\\
E_1^{\pm}=\omega_c\pm\sqrt{2}g;~|\Psi_1^{\pm}\rangle=\frac {1}{\sqrt{2}}|1,g,g\rangle\pm\frac {1}{2}(|0,e,g\rangle+|0,g,e\rangle),\\
E_2^{0}=2\omega_c;~|\Psi_2^0\rangle=\frac {1}{\sqrt{2}}|1,g,e\rangle-\frac {1}{\sqrt{2}}|1,e,g\rangle\nonumber\\
~\texttt{or}~~|\Psi_2^0\rangle=\frac {1}{\sqrt{3}}|2,g,g\rangle-\frac{2}{\sqrt{6}}|0,e,e\rangle,\\
E_2^{\pm}=2\omega_c\pm\sqrt{6}g;\nonumber\\
|\Psi_2^{\pm}\rangle=\frac {1}{\sqrt{6}}|0,e,e\rangle+\frac {1}{\sqrt{3}}|2,g,g\rangle\pm\frac{1}{2}(|1,e,g\rangle+|1,g,e\rangle),\\
E_3^{0}=3\omega_c;~|\Psi_3^0\rangle=\frac {1}{\sqrt{2}}|2,g,e\rangle-\frac {1}{\sqrt{2}}|2,e,g\rangle\nonumber\\
~\texttt{or}~~|\Psi_3^0\rangle=\frac {\sqrt{2}}{\sqrt{5}}|3,g,g\rangle-\frac{\sqrt{3}}{\sqrt{5}}|1,e,e\rangle,\\
E_3^{\pm}=3\omega_c\pm\sqrt{10}g;\nonumber\\
|\Psi_3^{\pm}\rangle=\frac {1}{\sqrt{5}}|1,e,e\rangle+\frac {\sqrt{3}}{\sqrt{10}}|3,g,g\rangle\pm\frac{1}{2}(|2,e,g\rangle+|2,g,e\rangle).
\end{gather}

In the case of driving the cavity by a classical laser field, the photon blockade will not occur, as displayed in Fig.~\ref{fig3}(b). The cavity photons still have coherent statistics, with the value of the two-order correlation function approximately equal to 1. The reason is that the absence of strong nonlinearity in the target system. The weak nonlinearity cannot prevent the system from absorbing the second and third photons when there is already a single photon in the cavity. We now consider exciting the target by the output field of source system rather than the external coherent field directly, and the two-photon blockade with two-photon bunching and three-photon antibunching can be obtained even though the target is in the weak coupling regime, as described by the gray area of Fig.~\ref{fig3}(a). The result is obtained by calculating numerically the master equation~(\ref{eq3}), and the two-photon blockade has potential for generating the entangled photon pairs. It is noteworthy that the scattered photons from the source system is in nature the non-classical light, i.e., quantum light, with the correlations $g^{(2)}, g^{(3)}<1$, as shown in Fig.~\ref{fig3}(d). But these photons do not exist in the form of photon pairs in the case of without atom-cavity coupling in the target or using the two-level system as a source directly, which is due to the absence of the strong two-photon correlation ($g^{(2)}<1$).  However, the two-photon blockade can happen in the atom-cavity coupled target system because the scattered photons from the source two-level emission themselves have quantum properties, making it difficult for the strong high-order photon correlations to occur in the target cavity. But the interaction between the cavity and atoms can enhance the photon correlations. Under competition of these two situations, the system induces the two-photon transition $(|\Psi_0\rangle\rightarrow|\Psi_1^\pm\rangle\rightarrow|\Psi_2^\pm\rangle)$, but the three-photon transition process is blockaded ( $|\Psi_2\rangle\nrightarrow|\Psi_3^\pm\rangle$), since the weak interaction is unable to make the system has a multiphoton transition from a low energy level to a higher energy level. Whereas this is no longer the case for using the two-level system as a source directly or for exciting the cavity QED system by the classical light directly.

Actually, the two-photon blockade has been reached in atom-cavity coupled systems excited by the classical light. However the essential requirement for achieving the effect is that these systems need to be in strong coupling regimes~\cite{ref20,ref19}. Here, our study based on the cascaded system can bypass the need to use technologically challenging strongly coupled cavities, and the influence of the counter-rotating terms of atom-cavity interaction on the photon statistics because exciting the target system with quantum light can enhance the non-classicality of the target~\cite{ref31-1}. The target system driven by the photon state with non-coherent statistics has stronger non-classical behavior than the one excited by classical light. Moreover, we also calculate the cavity photon correlations in same parameters when the target system contains one atom, as displayed in Fig.~\ref{fig3}(c). It is seen that the two-photon blockade in this system cannot be obtained. The reason for it is that the system consisting of a cavity coupled to an atom has weaker total interaction strength and nonlinearity than that coupled two atoms.
\begin{figure}
  \centering
  \includegraphics[width=8.5cm]{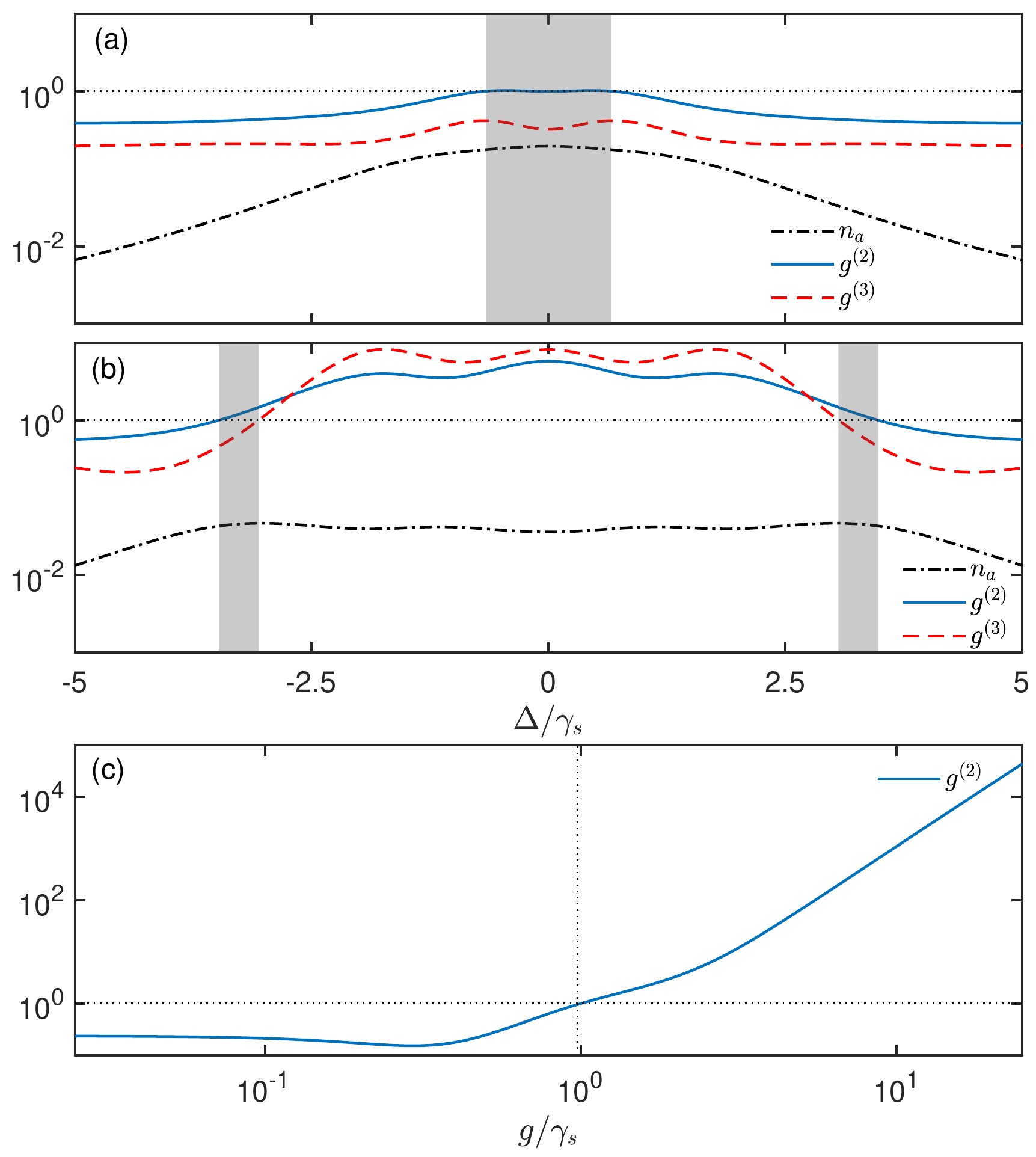}\\
  \caption{ The mean photon number $n_a$ (black dash-dotted curve), equal-time second-order photon correlation function $g^{(2)}$ (blue solid curve) and equal-time third-order photon correlation function $g^{(3)}$ (red dashed curve) of the cavity vs $\Delta_C$ for (a) $g/\gamma_s=1$ and (b) $g/\gamma_s=2.5$. (c) The equal-time second-order photon correlation function $g^{(2)}$ vs $g$ for $\Delta_C=0$. The other system parameters used here are the same as in Fig.~\ref{fig2}.}\label{fig5}
\end{figure}

In the preceding discussion, using the photon correlation functions to pinpoint the photon blockade effects, while customary, is not completely convincing, since the calculations involve very small denominators, i.e., average photon numbers. Therefore, we calculate numerically the photon-number probabilities $P_n$ and the fidelity of the $k$-photon truncation $F_k$ in the limit of steady-state. In Fig.~\ref{fig4}, we show the photon-number probabilities $P_n=|\langle n|\rho_{ss}\rangle|^2$ $(n=0,1,...)$ versus the cavity-laser frequency detuning $\Delta_C$, where $\rho_{ss}$ is the density operator of the steady-state. It is seen that the probabilities of one-photon (black solid line) and two-photon (red dotted line) states are nonzero, and the probability of three-photon state (green solid line) is basically zero. The results demonstrate that the two-photon blockade can occur in our proposal. The reason is that the presence of one-photon and two-photon Fock states blockade the generation of more photons in the target system~\cite{ref7}. However, the probabilities of generating one- and two-photon states are smaller than the probability of observing the zero-photon state because the statistical characteristics of quantum light make it difficult for the target system to have a large number of photons. This one can be seen in Fig.~\ref{fig3}(a), which shows very small values of average photon numbers. In order to observe the probabilities of the photon number states more clearly, we display the dependence of the ratio of two-photon state probability to one-photon state probability $R_{21}$ (blue dash-dotted line) on the values of $\Delta_C$, which has an obvious distribution in Fig.~\ref{fig4}. Moreover, to better understand the behavior of a two-photon blockade, we calculate the fidelity of the $k$-photon truncation, where the fidelity is defined as~\cite{ref7}
\begin{equation}\label{eq16}
F_k=\sum\limits_{n=0}^{k}P_n=\sum\limits_{n=0}^{k}|\langle n|\rho_{ss}\rangle|^2.
\end{equation}
From Fig.~\ref{fig4}, we see that $F_2\approx1$ and $F_1<1$ for the range of the probability of two-photon states, and the cavity field shows the sub-Poisson photon-number statistics. Although $F_1$ has a larger value with higher vacuum state population, caused by the quantum characteristics of the input field from source system, the result can still represent the two-photon blockade effect in our model.

Figure~\ref{fig5} shows that the mean photon number and photon correlations versus $\Delta_C$ for different coupling strengths between the cavity and atoms. By comparing Figs.~\ref{fig3}(a),\ref{fig5}(a) and~\ref{fig5}(b), we see that the maximum value of the second-order photon correlation decreases with the decrease of the coupling strength $g$. It can be predicted that the two-photon blockade will not occur when the value of $g$ continues to decrease. With increasing the coupling strength, $g^{(3)}>1$ can be seen for the small cavity-laser detunings in Figs.~\ref{fig5}(b), so the two-photon blockade effect can occur in the parameter regimes, as described by the gray areas. In Fig.~\ref{fig5}(c), we display the dependence of the second-order photon correlation on the coupling strength $g$. We see that the value of $g^{(2)}$ is smaller than 1 for $g/\gamma_s<0.9855$. In the range of $g/\gamma_s<0.9855$, the target system cannot observe the two-photon effect when the values of other parameters remain unchanged. The reason is that the weaker interaction makes it difficult to induce the multiphoton excitation in the target system.

\subsection{Two-photon blockade with atom-cavity detuning}
\begin{figure}
  \centering
  \includegraphics[width=8.5cm]{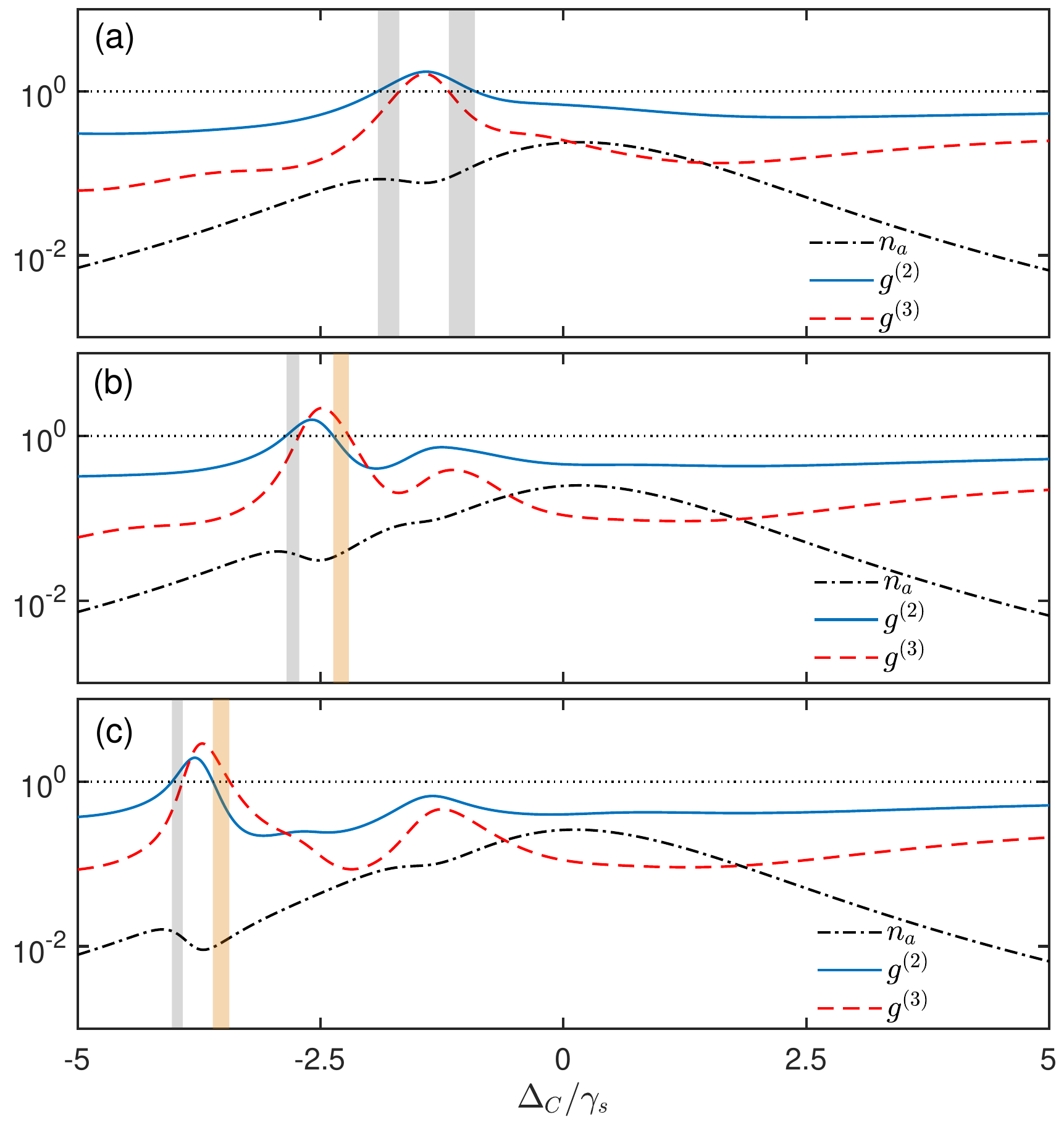}\\
  \caption{ The mean photon number $n_a$ (black dashed-dotted curve), equal-time second-order photon correlation function $g^{(2)}$ (blue solid curve) and equal-time third-order photon correlation function $g^{(3)}$ (red dashed curve) of the cavity versus $\Delta_C$. The system parameters used here are the same as in Fig.~\ref{fig2} except for $g/\gamma_s=0.9$, $\Delta_1/\gamma_s=\Delta_C/\gamma_s=1.25$. And $\Delta_2=\Delta_1$ (a), $\Delta_2/\gamma_s=\Delta_1/\gamma_s=1.25$ (b), and $\Delta_2/\gamma_s=\Delta_1/\gamma_s=2.5$ (c).}\label{fig6}
\end{figure}
We now consider that there is an out-of-resonance coupling between the cavity and atoms in the target system. We when assume $\omega_1=\omega_2=\omega\neq\omega_c$, and transform back to the laboratory frame, the dressed states and its eigenvalues of $H_t$ in three-photon space are given by
\begin{gather}\label{eq17}
E_1^{\pm}=\frac{\omega}{2}+\frac{\omega_c}{2}\pm\frac{\sqrt{8g^2+\omega^2-2\omega\omega_c+\omega_c^2}}{2};\nonumber\\
|\Psi_1^{\pm}\rangle=\frac {1}{\sqrt{2}}|1,g,g\rangle\pm\frac {1}{2}(|0,e,g\rangle+|0,g,e\rangle),
\end{gather}
\begin{gather}\label{eq18}
E_2^{\pm}=\omega+\omega_c\pm\frac{\sqrt{24g^2+\omega^2-2\omega\omega_c+\omega_c^2}}{2};\nonumber\\
|\Psi_2^{\pm}\rangle=\frac {1}{\sqrt{6}}|0,e,e\rangle+\frac {1}{\sqrt{3}}|2,g,g\rangle\pm\frac{1}{2}(|1,e,g\rangle+|1,g,e\rangle),
\end{gather}
\begin{gather}\label{eq19}
E_3^{\pm}=\frac{3\omega}{2}+\frac{3\omega_c}{2}\pm\frac{\sqrt{40g^2+\omega^2-2\omega\omega_c+\omega_c^2}}{2};\nonumber\\
|\Psi_3^{\pm}\rangle=\frac {1}{\sqrt{5}}|1,e,e\rangle+\frac {\sqrt{3}}{\sqrt{10}}|3,g,g\rangle\pm\frac{1}{2}(|2,e,g\rangle+|2,g,e\rangle).
\end{gather}
By calculating numerically Eq.~(\ref{eq3}), the dependences of the mean photon number and photon correlation functions on the cavity-laser frequency detuning $\Delta_C$ are shown in Fig.~\ref{fig6}. From Fig.~\ref{fig6}(a), it is seen that the two-photon blockade can be obtained when the atom-cavity detunings are considered, as displayed by the gray areas. Note that this two-photon blockade effect cannot happen when the cavity coupled to two identical atoms in the same coupling regime, which has been discussed in the above section. However, the reason for the behavior that happened here is that the detunings between the cavity and atoms increase the nonlinearity of the target system, making it possible to realize the two-photon transition between the cavity and atoms $(|\Psi_0\rangle\rightarrow|\Psi_1^\pm\rangle\rightarrow|\Psi_2^\pm\rangle)$.

Moreover, from Figs.~\ref{fig6}(b) and~\ref{fig6}(c), we find that the two-photon blockade effect can also occur when the target cavity weakly coupled to two atoms with different transition frequencies. It is noteworthy that the yellow areas of Figs.~\ref{fig6}(b) and~\ref{fig6}(c) correspond to the two-photon correlation being suppressed and the three-photon correlation being enhanced via the high-order transition processes. Here the result shows that the occurrence of the unconventional PB effect in target system due to the presence of the inhomogeneous broadening of atoms~\cite{ref12}. Comparing Figs.~\ref{fig6}(b) and~\ref{fig6}(c), we find that, the range of frequency for realizing the unconventional PB in a large homogeneous broadening is wider than that of a small homogeneous broadening. However, the result is reversed for realizing the two-photon blockade in the same parameters.

\section{Discussion}
In this section, we focus on the discussion of the applications and experimental realization of our proposal. We know that the multiphoton blockade has been observed in various systems~\cite{ref7,ref17,ref17-1,ref18,ref20,ref19} and the single-photon blockade effect has already been reached experimentally~\cite{ref2-1,ref14,ref3}. The main request for achieving these effects is the presence of strong nonlinearities. However, it is still difficult to achieve the strong nonlinearities in many systems because generation of strong nonlinearity needs strong nonlinear susceptibility, a strong coupling regime or a high-quality cavity. With the progress of technology, the light-matter interaction in the strong coupling regime has been studied in a variety of cavity QED systems~\cite{ref32,ref33}, and even the ultrastrong coupling between light and matter can also be achieved in superconductor and solid-state semiconductor systems~\cite{ref34,ref35}. The PB effect under a weak coupling regime is still a field worth studying, since comparison with a high-quality cavity, a weak coupling regime or a bad cavity system are easily obtained in experiments. Moreover, for the weak coupling regime, we can bypass the influences of the counter-rotating terms of atom-cavity interaction on the photon-number statistics.

It is already known that, in the weak coupling or bad cavity regime, the PB can be obtained with great difficulty in a system driven by classical coherent light. Thus we consider driving the target system by non-classical light from a source system rather than classical light from an external coherent light field, since exciting the target system with a photon state having a non-coherent statistics can enhance the non-classicality of the system~\cite{ref31-1}. This process can occur in a cascaded quantum system. For such a quantum system, we investigate the multiphoton blockade and unconventional PB effects as bypassing the need to use technologically challenging strongly coupled cavities. The two-photon blockade and unconventional PB can be obtained even when the target system is in the weak coupling or bad cavity regime. The research extends the regime of the generation of two-photon blockade, making it possible for the PB effect to occur in various systems. The PB can be used to indicating the non-classicality of the systems~\cite{ref36-1}, and is useful for generating the non-classical photon pairs, which could be used as the ideal entangled photon sources. Moreover we also face many challenges in the research of the two-photon blockade. In practice, only a small number of photons from the source can be used to drive the target cavity in the cascaded system, as compared with the case of producing the multiphoton blockade in a strong atom-cavity coupling system driven by a classical light field ~\cite{ref20,ref19}. From Figs.~\ref{fig3}(a) and~\ref{fig4}, it is seen that the average photon number $n_a\approx0.1681$ and the probability of two-photon state $P_2\approx0.01945$ when $\Delta_C/\gamma_s=0$. Thus we can see that the system needs to produce about $8.6$ (i.e., $n_a/P_2\approx8.6$) cavity photons to obtain a single photon pair in that case. The probability of producing the photon pair in this system is much lower than that of other systems driven by classical light (e.g., the Kerr-type system~\cite{ref7}). The improvement of the scattered photons that can be scanned onto the target is necessary for harnessing the two-photon blockade effect.

The target system is driven by the output field from source system, and the two subsystems can be coupled through a Lindblad term in the master equation. The coupling strength is dependent on the decay rates of two subsystems, but the decays will also affect the dissipation and evolution of the coupled system. Therefore, we need to use appropriate parameter regimes to achieve the two-photon blockade before the coupled system is dissipated. From Fig.~\ref{fig2}, we see that the atoms and cavity can undergo many excitation number conversions before the decoherence of system. Thus our proposal is effective in the coherent time. Furthermore, we consider the experimental realization of our proposal. Based on these articles~\cite{ref36-2,ref36-3,ref36-4,ref36-5}, we construct the target system that consists of a photonic crystal cavity with two coupled quantum dots (QDs). The system allows us to control the resonance (or detuning) between the cavity and quantum dots with a laser. Here the cavity field decay rate is $\kappa/2\pi=16$~GHz (i.e., the lifetime $\tau_c\sim10$~ps), corresponding to a quality factor $Q=10^4$. The QDs have the spontaneous emission rate of $\gamma/2\pi=1.2$~GHz (i.e., the lifetime $\tau\sim133$~ps), and the coupling strength of $g/2\pi=4$~GHz. For the source system, based on these experimental articles~\cite{ref37,ref38,ref39}, we consider a configuration consisting of individual self-assembled (In, Ga) As/GaAs QD embedded in a microcavity. The system is maintained at low temperature in a continuous-flow cryostat, and a polarization-maintaining single-mode optical fiber is brought close to the system edge. Then, we prepare a coherent laser field, made of the coherent pump field and vacuum field in a unitary mixer, coupling into the waveguide mode of the cavity through the fiber. Here, the vacuum field, as one of the inputs of the source, is used to prevent the target system from also being driven by the coherent pump field~\cite{ref27,ref28,ref29,ref31-1}. Some fraction of the QDs coupled to the single-mode cavity, and photons can be scattered from the cavity. These scattered photons are tunable non-classical lights, which could be used to excite the target system. Here, the output light of the source system drives the target cavity via a waveguide supporting only a right-propagating mode~\cite{ref40,ref41}. In particular, for the cascaded coupling system, we do not need to consider the back action from the target and the time delays between source and target systems. The emitted photons from the target system are directed into the setups to measure photon correlations $g^{(2)}$ and $g^{(3)}$~\cite{ref42,ref43,ref19}.

\section{Conclusion}
In summary, we have studied the photon statistics of cavity QED system that consists of a cavity coupled to two two-level atoms in the weak coupling regime. Here, the cavity is excited by quantum light from a source system, and the source is made of a two-level atom driven by classical laser field. The two-photon blockade with two photon bunching and three photon antibunching can be observed in our model even when the strong system dissipation is included. We have also investigated the photon number statistics in the truncated Hilbert space in the same parameter regimes, whose results also demonstrated that the two-photon blockade effect can occur in our model. Moreover, this two-photon blockade effect can also be observed in a weaker coupling regime when the cavity-atom detunings are considered. In particular, in this regime, we have also found the unconventional PB effect when the cavity was weakly coupled to two nonidentical atoms, where two-photon correlation was suppressed and three-photon correlation was enhanced. This work studied quantum statistical characters of cavity QED system in cascaded quantum system, and showed the photon blockade effect in the weak coupling regime, which should advance the development of generating photon pairs and have potential applications in quantum information science.

\begin{acknowledgments}
This work is supported by National Key Research and Development Program of China (No. 2016YFA0301203); National Science Foundation of China (NSFC) (11374116, 11574104, 11375067).
\end{acknowledgments}

\end{document}